\documentclass[pra,floatfix,showpacs,twocolumn]{revtex4}
\usepackage{graphicx}
\usepackage{amssymb,amsfonts,amsmath}
\usepackage{color}
\usepackage{ulem}

\usepackage[T1]{fontenc}

\newcommand{\nc}{\newcommand}
\nc{\rnc}{\renewcommand}
\nc{\nn}{\nonumber}

\nc{\g}{\gamma}
\nc{\om}{\omega}
\rnc{\b}{\beta}
\rnc{\th}{\theta}

\rnc{\[}{\left[}
\rnc{\]}{\right]}
\rnc{\(}{\left(}
\rnc{\)}{\right)}

\newcommand{\bra}{\langle}
\newcommand{\ket}{\rangle}
\nc{\vac}{|0\ket}
\nc{\vvac}{\bra0|}

\nc{\cd}{\cdots}
\nc{\sm}[2]{\sum_{#1=1}^{#2}}

\nc{\red}{\textcolor{red}}
\nc{\sred}[1]{\textcolor{red}{\sout{#1}}}

\nc\hp{\hat{\psi}}
\nc\hpd{\hat{\psi}^\dagger}

\nc\qs{\psi_\text{QS}}
\nc{\mfinf}{\psi_\text{MF}^{(\infty)}}
\nc{\phiinf}{\phi_\text{MF}^{(\infty)}}
\nc{\thinf}{\th_\text{MF}^{(\infty)}}
\nc{\mfperi}{\psi_\text{MF}}
\nc{\phiperi}{\phi_\text{MF}}
\nc{\thperi}{\th_\text{MF}}
\nc{\fsn}{f_{\text{sn}}}
\nc{\gsn}{g_{\text{sn}}}
\nc{\hsn}{h_{\text{sn}}}

\begin{document}
\title{Recurrence Time in the Quantum Dynamics of the 1D Bose Gas}

\author{Eriko Kaminishi$^1$}
\author{Jun Sato$^2$}
\author{Tetsuo Deguchi$^3$}

\affiliation{$^1$Department of Physics, 
%Graduate School of Science, 
University of Tokyo, \\ 
7-3-1 Hongo, Bunkyo-ku, Tokyo 113-0033, Japan} 
\affiliation{$^2$
Research Center for Advanced Science and Technology, University of Tokyo,
4-6-1 Komaba, Meguro-ku, Tokyo 153-8904, Japan
}
\affiliation{$^3$ Department of Physics, 
Graduate School of Humanities and Sciences, 
Ochanomizu University, 2-1-1 Ohtsuka, Bunkyo-ku, Tokyo 112-8610, Japan}
\date{\today}
% ___________________________________________________________________________________________

\begin{abstract} 
Recurrence time is evaluated for some initial quantum states in the one-dimensional Bose gas with repulsive short-range interactions. In the relatively strong and weak coupling cases some different types of initial states show almost complete recurrence and the estimates of recurrence time are proportional  to some powers of the system size at least in some range of the system size. They are much longer than in the case of free particles such as 100 times.  In the free-bosonic and free-fermionic regimes  we evaluate the recurrence time rigorously, which is proportional to the square of the system size. The estimate of recurrence time is given by the order of ten milliseconds in the corresponding experimental systems of cold atoms trapped in one dimension of ten micrometers in length. It is much shorter than the estimate in  a generic quantum many-body system, which may be as long as the age of the universe. 
\end{abstract} 
\maketitle

\section{Introduction}
Recurrence is one of the fundamental concepts not only in 
classical mechanics but also in quantum statistical mechanics \cite{Hobson}. 
It has become quite attractive to study  theoretically recurrent phenomena 
in isolated quantum systems, due to recent experiments of cold atomic systems confined in one dimension \cite{Ketterle,Esslinger,Kinoshita,dark-soliton,prethermal}.   
They have created a huge motivation for studying 
fundamental aspects of quantum statistical mechanics: Equilibration or relaxation of isolated quantum many-body systems 
\cite{Rigol} and ergodic theorems \cite{vonNeuman} 
in quantum statistical mechanics from the  viewpoint of typicality 
\cite{Tasaki,Lebowitz,Reimann,Sugita}.  
Furthermore, the dynamics of isolated quantum many-body systems 
in one dimension has been extensively studied by both experiments 
\cite{Kinoshita,dark-soliton,prethermal,I.Bloch2002-2010,I.Bloch2004,I.Bloch2012} 
and theories \cite{Rigol,Polkovnikov,Yukalov, Andrei,Konik,Ikeda,SKKD1,Spontaneous, Kaminishi2014}, 
in particular, associated with quantum quenches in quantum spin systems    
\cite{McCoy1970,SeiSuzuki,Mossel2010,Igloi-Rieger,Rigol2011,Calabrese-Essler} 
and in conformal field theories (CFT) \cite{Calabrese-Cardy,Sotiriadis,Cardy}.   

Almost-periodic  quantum dynamical phenomena are observed in recent experiments.
Oscillating behavior was observed in experiments of cold atomic gases in one dimension \cite{Kinoshita}. The system is close to the integrable system of the one-dimensional (1D) interacting bosons with the delta-function potentials, which we call the 1D Bose gas with the delta-function interactions. 
Quantum collapse and revival in a one-atom maser were experimentally demonstrated, which had been studied in the Jaynes-Cummings model \cite{QCR,Collapse}. 
Oscillating behavior of the BEC such as breathing mode \cite{breathing} and scissors mode \cite{scissors} was observed. 
Almost full revivals for atoms were observed experimentally in optical lattices 
\cite{I.Bloch2002-2010}.

Although every generic isolated quantum system is almost periodic, 
a recurrence phenomenon is usually not observed even in the numerical simulation unless we choose the system and the initial state quite properly. It is proven that  for a quantum state given by the superposition of a discrete set of energy eigenstates the time evolution is almost periodic,  i.e., a quantum analogue of Poincare's recurrence theorem \cite{Ono,QRT,Percival,Hobson}. It is also demonstrated that quantum systems with time-periodic Hamiltonians are almost periodic \cite{Hogg}. However,  recurrence time  is very long for generic quantum systems with incommensurable energy levels. Typically it is proportional to the exponential of the number of the eigenstates in the superposition of a discrete set of eigenstates \cite{Peres}. 
It therefore may be  extremely long such as long as the age of the universe.  
In some spin system with long-range interactions it is analytically shown that the recurrence time is proportional to the exponential of the number of particles \cite{Emch,Radin,Kastner}. It is much shorter than the exponential of the number of superposed eigenstates, but is still very long.

In this paper, we present concrete examples of recurrence for some types of  quantum states in the 1D Bose gas with the delta-function interactions,  
and we evaluate numerically the recurrence time for them.   
Let us call the squared amplitude between an initial state and the time-evolved state the {\it fidelity}. We numerically determine the recurrence time by the shortest interval of time in which the value of the fidelity returns to a value larger than 0.9 for the first time in the time evolution after the initial time.  
We observe periodic patterns in the time evolution of the fidelity for different values of the interaction strength. We also show numerically the recurrence of the density profile for the quantum states. Here we expect that the density profile can be measured  in experiments, while it is not easy to measure the fidelity.

The recurrence time depends on the initial state. In this paper, we consider two types of initial states. The superposition of one-hole excitations, and that of two-hole excitations:  
They are important in the low-lying excitation spectrum. Here we remark that some superposition of one-hole excitations leads to  a quantum dark soliton \cite{quantum-soliton}. The estimates of recurrence time are proportional to some powers of the system size at least up to certain sizes for some finite nonzero values of the coupling constant $\gamma$ which are large such as $\gamma=10^2$ or small such as $\gamma=10^{-2}$. Here we shall define the coupling constant $\gamma$ in section II. 
In the case of infinite or zero interaction strength, we derive rigorously the recurrence time for any given state and it is proportional to the square of the system size.  For intermediate values of the coupling constant $\gamma$ satisfying $ 10 ^{-2} \ll \gamma \ll 10^2$, however, it seems that the fidelity shows neither any periodic pattern as a function of time nor any recovery to a value larger than 0.9 within our range of computational  time, 
and hence we do not evaluate recurrence time for them.     

We give numerical estimates of recurrence time in experimental systems of the 1D Bose gas with the delta-function interactions. It is given by the order of ten milliseconds if the quantum system is realized in cold atomic gases trapped in one dimension of ten micrometers in length. It is much shorter than the age of the universe.
We suggest that it requires a high degree of isolation from the environment 
to observe recurrence phenomena experimentally.    
Moreover, it is not clear how to observe them in such an ``almost isolated'' quantum system of cold atoms with a long coherence time. 
However, the estimated recurrence time in the paper,  which we expect can be shorter than the  coherence time, should be nontrivial and motivate further studies.  

Recurrence in the 1D Bose gas  with the delta-function interactions should be important in many aspects of the quantum dynamics of many-body systems such as in Ref.  \cite{Quan} where the recurrence of fidelity is investigated numerically. The 1D Bose gas has nonlinear excitation modes, which play a key role in the nonlinear TL liquid \cite{Imambekov}.      
The low-lying excitations of the 1D Bose gas are well approximated by the linear bosonic modes, which are described in terms of the Tomonaga-Luttinger (TL) liquid or the CFT with central charge $c=1$ \cite{Korepin}. For some superposition of excited states in a linear mode, the recurrence time or the time period for revival may be proportional to the system size,  as argued in CFT \cite{Cardy}.

The contents of the paper consist of the following. In Sec. II we give the Hamiltonian of the 1D Bose gas with the delta-function interactions, which we call the Lieb-Liniger (LL) model \cite{Lieb-Liniger}. We give notation of the Bethe ansatz, and introduce two types of quantum states, the state given by the sum over one-hole excitations, 
the sum over two-hole excitations. 
We also introduce dimensionless time variable $t$, which is useful to show the system-size dependence for the estimates of recurrence time.  
In Sec. III we evaluate recurrence time rigorously for the 1D Bose gas with  the delta-function interactions in the free-fermionic and the free-bosonic regimes. In Sec. IV we show that 
in the case of relatively strong or weak coupling,  the fidelity of a state in the two types shows almost complete recurrence: it returns to a value larger than 0.9. We evaluate the recurrence time for the states given by the sum over one-hole excitations and the sum of two-hole excitations. 
We show how it increases with respect to the system size.  
With the determinant formula of the form factors \cite{Slavnov1989},
we evaluate the density distribution of the 1D Bose gas with delta-function interactions.  
We confirm that when the density distribution returns to the initial form 
the fidelity also becomes close to 1.0 as far as in the sum over one-hole excitations. 
We thus suggest that if the density profile recurs, the fidelity becomes close to 1.0 practically for such states with small particle numbers  as we have investigated in the present paper.  
In Sec. V we give numerical estimates of recurrence time for some experimental systems. Finally in Sec. VI we give concluding remarks.

%%%%%%%%%%%%%%%%%%%%%%%%%%%%%%%%%%%%%%%%%%%%
\section{The Lieb-Liniger Model }

\subsection{Lieb-Liniger Hamiltonian}

Let us introduce the Hamiltonian of the 1D Bose gas with the delta-function interactions, 
which we call the LL model \cite{Lieb-Liniger}, as follows. 
\begin{align}
{H}_{\text{LL}} 
= -\frac{{\hbar}^2}{2m} \sum_{j=1}^{N} {\frac {\partial^2} {\partial q_j^2}}
+  2g \sum_{j < k}^{N} \delta(q_j-q_k) . \label{eq:LLH}
\end{align}
\par \noindent 
Here, $N$ bosons with mass $m$ are interacting through the delta-function potentials with the coupling constant $g$. We now denote the time variable by $\tau$. The Schr{$\ddot{\rm o}$}dinger equation at time $\tau$ is given by 
\begin{align}
i\hbar\frac{\partial}{{\partial}{\tau}} |\Phi ({\tau}) \rangle = {\cal H}_{\text{LL}} |\Phi (\tau)\rangle. \label{eq:SE}
\end{align}
We assume that the wavefunctions satisfy the periodic boundary conditions  
of the system size $L$. Here, the coordinates $q_j$ satisfy {$0 \leq q_j \leq L$. 
We introduce  the coupling constant $c$ by  $g ={ {\hbar}^2 c}/{2m}$}.  
Hereafter, we consider the repulsive interaction: $ c > 0 $. 

It is known that the bulk quantities of the LL model are characterized by the parameter $\gamma = c/n$ and $N$, where $n = N/L$ is the particle density \cite{Lieb-Liniger}. %\sout{(see also Appendix A)}. 
We define  dimensionless coordinate variables $x_j$ by 
\begin{align}
q_j =x_j/n , \quad (0 \leq x_j \leq N)
\end{align}
and dimensionless time variable $t$ by    
\begin{align}
\tau={2mt}/{{n}^2\hbar} \, .    
\label{tex}
\end{align}
We express the Hamiltonian (\ref{eq:LLH}) and the Schr{$\ddot{\rm o}$}dinger equation (\ref{eq:SE}), respectively,  as 
\begin{align}
{H'}_{\text{LL}} 
= - \sum_{j=1}^{N} {\frac {\partial^2} {\partial {x}_j^2}}
+ 2{\gamma} \sum_{j < k}^{N} \delta(x_j-x_k) ,  
\label{hprime}
\end{align}
and 
\begin{align}
i \frac{\partial}{{\partial}t} |\Phi'(t)\rangle = {\cal H'}_{\text{LL}} |\Phi'(t)\rangle . 
\label{sprime}
\end{align}
Here the symbol $|\Phi'(t)\rangle$ denotes 
$|\Phi'(t)\rangle = |\Phi (\tau)\rangle$. 

Hereafter in the paper we mainly employ the dimensionless time variable $t$   
and make use of eqs. (\ref{hprime}) and (\ref{sprime}) rather than eqs. 
(\ref{eq:LLH}) and (\ref{eq:SE}), respectively. 
We shall show that the dimensionless time variable 
is useful to express the system size dependence of  
recurrence time, explicitly. 
However,  we return to the original time variable $\tau$ when we estimate the recurrence time for experimental systems in Sec. V.

% \sout{\textcolor{red}{The effect of trap potential, heating and  other relevant processes becomes relevant only for long time scales, so we ignore there effects.
% As is shown later, the recurrence time is relatively short for weak and strong coupling regime, and hence this approximation would be justified.}}

%%%%%%%%%%%%%%%%%%%%%%%%%%%%%%%%%%%%%%%%%%%%%%%
\subsection{The Bethe Ansatz Equations }

In the LL model, the Bethe ansatz offers an exact eigenstate 
with an exact energy eigenvalue for a given set of quasi-momenta 
$k_1, k_2, \ldots, k_N$ satisfying 
the Bethe ansatz equations for $j=1, 2, \ldots, N$.
\begin{align}
k_j N = 2 \pi I_j - 2 \sum_{\ell \ne j}^{N}
\arctan \left(\frac {k_j - k_{\ell}} {\gamma} \right). 
\label{BAE} 
\end{align}
Here $I_j$'s are integers for odd $N$ and half-odd integers for even $N$. 
We call them the Bethe quantum numbers. 

The total momentum $P$ of the system of $N$ bosons 
is given by the sum of all $k_j$'s: $P  = \sm{j}{N}k_j$. 
It follows from (\ref{BAE}) that we have 
\begin{equation}
P=\frac {2 \pi} N \sum_{j=1}^{N} I_j . \label{eq:momentum} 
\end{equation}
The energy eigenvalue $E$ of the Hamiltonian (\ref{hprime}) is 
expressed in terms of the quasi-momenta as 
\begin{equation}
E = \sm{j}{N}k_j^2 . \label{eq:energy}
\end{equation} 

For the ground state of $N$ bosons, 
the quantum numbers $I_j$ are given  by 
\begin{equation} 
I_j= j -(N+1)/2 \, , \quad \mbox{for} \quad j = 1, 2, \ldots, N \, . 
\end{equation} 
%

%%%%%%%%%%%%%%%%%%%%%%%%%%%%%%%%%%%%%%%%%%%%%%%%%%
\subsection{Superposition of One-Hole Excitations}

Superposing Lieb's type II excitations \cite{Lieb-Liniger}, 
i.e. one-hole excitations, we construct a quantum state 
with an initially localized density profile.  
We remark that it coincides with the amplitude profile of a dark-soliton solution 
of the Gross-Pitaevskii equation \cite{quantum-soliton}. 
In the type II branch, for each integer $p$ in the set $\{0, 1, \ldots, N-1\}$, 
we consider the one-hole excitation of $N$ particles, and   
the total momentum $P$ is given by $P=2 \pi p/{N}$.  
We denote the normalized Bethe eigenstate of $N$ particles 
with total momentum $P$  by $|P, N \ket$. 
The Bethe quantum numbers for the Bethe eigenstate $|P, N \ket$ are given by
\begin{equation} 
I_j  =  
\left\{ \begin{array}{c} 
 j-{\frac {N+1} 2}  \quad \mbox{for} \, \,  j = 1, 2, \ldots,  N-p, \\   
 j-{\frac {N+1} 2}+1 \quad \mbox{for} \, \,  j = N-p+1, \ldots, N. 
\end{array} \right. 
\end{equation}

The hole of the quantum state $|P, N \ket$ is located between the integers $I_{N-p}$ and $I_{N-p+1}$.  Here we have  $I_{N-p}=(N-1)/2 -p$ and $I_{N-p+1}= (N-1)/2-p +2$ and 
the difference is given by 2, not by 1: The integer $I_{HL}=(N-1)/2 -p+1$ is not occupied 
and gives a hole. Furthermore, we have a particle at $I_N=(N+1)/2$. 
We denote it also by $I_{\rm PT}$. By making use of eq. (\ref{eq:momentum}) 
the momentum $P$ is given by the difference between integers $I_{\rm PT}$ 
and $I_{\rm HL}$: 
\begin{equation} 
P= 2 \pi (I_{\rm PT} - I_{\rm HL})/L = 2 \pi p/L.     
\end{equation}  

For each integer $q$ satisfying $0 \le q \le N-1$ 
we define the coordinate state $|X \ket$ for $0\leq X\leq N$
by the discrete Fourier transform: 
\begin{align}
| X \ket := \frac 1 {\sqrt{N}} \sum_{p=0}^{N-1} 
\exp(- 2 \pi i p X/N) \, | P, N \ket \, . 
\label{eq:XN}
\end{align}

We define the quantum state at time $t$, $| X (t) \ket$, by 
$| X (t) \ket := \exp(-i{H'}_{\text{LL}}t) | X \ket$.
Through formula (\ref{eq:energy}) we numerically obtain all the energy eigenvalues $E_p$ of one-hole excitations  $|P \ket$s' in the type-II branch.  
We can perform the time evolution of the quantum states 
for quite a long time.  We recall that the fidelity of the  quantum state 
$| X (t) \ket$ is given by the squared overlap between the initial state $| X(0) \ket$ 
and the time-evolved state $|X(t) \ket$ at time $t$ as 
$F(t) :={ \left| \bra X (t) | X (0) \ket \right| }^2$.

We remark that the quantum states $| X (0) \ket$ consisting of the type-II excitations are  important and physically relevant as the initial states. 
In fact,  the type-II excitations (the one-hole excitations)  
 correspond to the lowest energy eigenstates with given angular momenta, i.e. the Yrast states \cite{Yrast}, and we expect that they are stable at low temperature.

%%%%%%%%%%%%%%%%%%%%%%%%%%%%%%%%%%%%%%%

\subsection{Two-Hole Excitations }
 
Let us consider a Bethe eigenstate consisting of $N$ particles  
with integers $p_1$ and $p_2$ for two holes. 
We call it the two-hole excitation with integers  $p_1$ and $p_2$,  and 
denote it by $|p_1, p_2; N \rangle $. 
We assume that the integers satisfy conditions $1 < p_2 \le p_1 \le N$. 
They correspond to momenta $P_1$ and $P_2$ 
by  $P_1= 2 \pi p_1/L$ and $P_2= 2 \pi p_2/L$, respectively. 
The total momentum of the eigenvector $|p_1, p_2; N \rangle $ is given by the sum: 
\begin{equation} 
P_1+P_2= 2 \pi (p_1 +p_2)/L.  
\end{equation} 

The Bethe quantum numbers of  eigenstates $|p_1, p_2; N \ket$ for integers $p_1$ and $p_2$ satisfying  $1 < p_2 \le  p_1 \le N$ are given by
\begin{equation} 
I_j  =  
\left\{ \begin{array}{cc} 
j-{\frac {N+1} 2} & \mbox{for} \, \,  j = 1, 2, \ldots,  N-p_1, \\   
j-{\frac {N+1} 2}+1 & \mbox{for} \, \,  j = N-p_1 +1 , \ldots,  N-p_2, \\   
j-{\frac {N+1} 2}+2 & \mbox{for} \, \,  j = N-p_2+1, \ldots, N. 
\end{array} \right. 
\end{equation}
The quantum numbers of the two holes  in the eigenstate $|p_1, p_2; N \ket$ are located 
between the Bethe quantum numbers  $I_{N-p_1}$ and $I_{N-p_1+1}$, 
and  between the Bethe quantum numbers  $I_{N-p_2}$ and $I_{N-p_2+1}$, respectively. 
They are given by $I_{\rm HL1} = (N-1)/2- p_1 + 1$ and 
$I_{\rm HL2} = (N-1)/2 - p_2 +2$, respectively.   
Here, we have particles of $I_{\rm PT1}= (N-1)/2 +1$ and  $I_{\rm PT2}=(N-1)/2 +2$, 
respectively. 

We consider a quantum state which is given by the sum over two-hole excitations for $p_1 = 2, 3, \ldots, N$ 
with $p_2$ being fixed as $p_ 2= 2$,
\begin{eqnarray}
&& | X, p_2=2  \ket 
\nonumber \\ 
& & :=  \frac 1 {\sqrt{N-1}} 
\sum_{p_1=2}^{N} 
\exp(- 2 \pi i p_1 X/N) \, |p_1, p_2=2; N \ket . 
\nonumber \\ 
%| P, N \ket .
\label{x2}
\end{eqnarray}
We call it the sum of two-hole excitations.

%%%%%%%%%%%%%%%%%%%%%%%%%%%%%%%%%%%%%%%%%%%%%%%%
%
% Sec 3
%
\section{Free-Bosonic and Free-Fermionic Regimes}

\subsection{Derivation of Recurrence Time}

We now evaluate rigorously the recurrence time  
in the free-bosonic and free-fermionic regimes, 
where we have $\gamma=0$ and $\infty$, respectively. 

For an illustration, let us consider  the quantum state which is given by the sum 
over one-hole excited states in the branch.  
In the free-fermionic regime we consider one-hole excitation $|P, N \ket$ 
for each integer $p$ satisfying $0 \le p \le N-1$,  
in which we have a particle at $I_{\rm PT}=(N+1)/2$
 and a hole at $I_{\rm HL}= (N+1)/2-p$;    
in the free bosonic regime we have 
$p$ particles at $I=1$ and $N-p$ particles at $I=0$.   

Let us express the difference between the one-hole excited energy $E_p$ 
and the ground state energy $E_g$ as   
\begin{equation} 
E_p - E_g =   (2 \pi/{N})^2  e_p . 
\end{equation} 
Here,  $e_p$ is given by an integer. 
We have $e_p =\{(N+1)/2\}^2-\{(N+1)/2-p\}^2$ in the free-fermionic regime,
and $e_p = p$ in the free-bosonic regime. 

The fidelity at recurrence time $T$ is given by 
\begin{align}
F(T) =\frac 1 {N^2} \Big{|} \sum_{p=1}^{N-1} 
\exp{\{i\( {\frac{2\pi}{{N}}}\)^2e_{p} T \}}\Big{|}^2 .
\label{fidelity2}
\end{align} 
It follows from the condition, $F(T)=1$, that we have the recurrence time as follows. 
\begin{align}
T = \frac{ {{N}}^2 }{ 2{\pi}G } . 
\label{larget}
\end{align}
Here, $G$ is the greatest common divisor among the integers in the set 
$\{e_1,  e_2, ... , e_{N-1}\}$.   
The expression of recurrence time  (\ref{larget}) is exact. 
If we write eq.(\ref{larget}) in the dimensionful form, it is
\begin{align}
\tau_{\rm rec}={\frac{m{L}^2}{\pi \hbar G}} . 
\label{texperiment}
\end{align}

For various other quantum states, the recurrence time is given by 
the same formula (\ref{larget}) in  the case of $\gamma=0$ or 
$\gamma=\infty$.  
Let us consider a given Bethe ansatz eigenstate $| \{ k_j \} \rangle$ 
of $N$ particles with pseudo-momenta $k_1 , k_2, \ldots, k_N$.  
The energy difference between 
the excited state from the ground state with pseudo-momenta 
$k_1^{(g)},   k_2^{(g)}, \ldots, k_N^{(g)}$ is given by  
\begin{equation} 
\Delta E = \sum_{j=1}^{N} k_j^2 - \sum_{j=1}^{N} \left( k_j^{(g)} \right)^2 \, . 
\end{equation}   
We express it as 
\begin{equation} 
\Delta E  = \left( \frac {2 \pi} L \right)^2 e_{\rm ex}  .  
\end{equation} 
Then, $e_{\rm ex}$ is  always  given by an integer 
in the free-bosonic or free-fermionic regimes.
Therefore, for a quantum state given by the sum over several excited states 
the recurrence time  is given by 
(\ref{larget}) in dimensionless unit of time and by 
 (\ref{texperiment})  in dimensionful unit of time. 

 When the fidelity returns to 1 at time $t=T$,  
the state returns to the initial one 
except for a relative phase factor,  and all the physical quantities 
take the same values as in the initial state.

Recurrence time in the free-fermionic regime has even-odd dependence on 
the number of particles $N$. We can prove that 
recurrence time in the free-fermionic regime with odd $N$ 
is equal to that in the free-bosonic regime with the same $N$, 
while the recurrence time in the free-fermionic regime with even $N$ 
is half of that in the free bosonic regime with the same $N$.

\begin{figure}[h!]
\includegraphics[width=8cm]{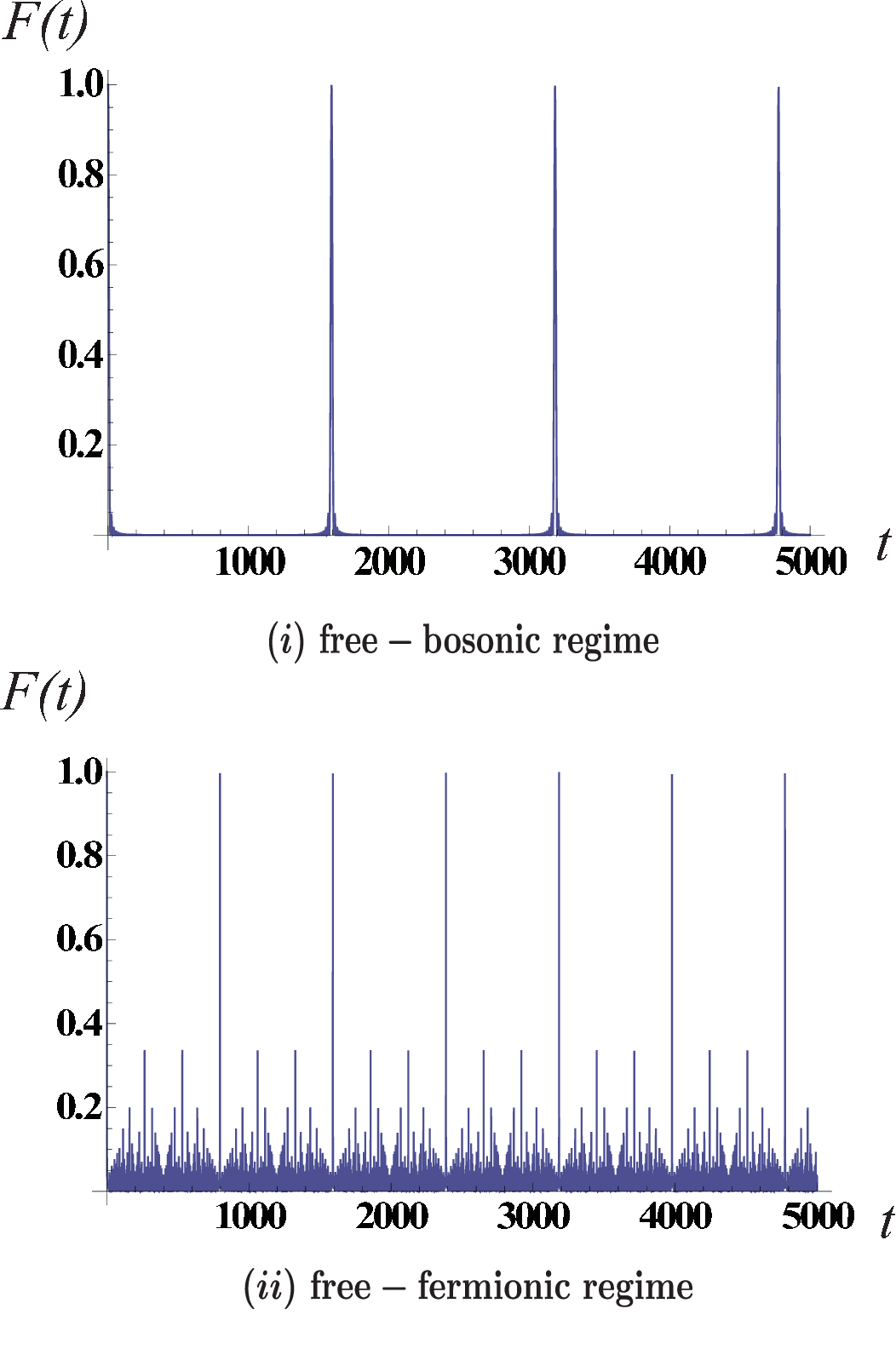}
\caption{(Color online) Complete recurrence: 
Fidelity of the superposition of the type II excitations,  
$F(t) = |\bra X(t) | X(0) \ket |^2 $, versus time $t$. 
In the upper panel: Free-bosonic regime;  
In the lower panel:  Free-fermionic regime,  
both for $N=1000$. }
\label{free}
\end{figure}

The rigorous results are confirmed and illustrated 
by numerical calculations.
In Fig. \ref{free} the time evolution of the fidelity for the state given by 
the superposition of the type II excitations,  
$F(t) = |\bra X(t) | X(0) \ket |^2 $, is plotted against time $t$. 
The fidelity returns to 1.0 completely and periodically in time.

%%%%%%%%%%%%%%%%%%%%% figure1 %%%%%%%%%%%%%%%%%%%%%%%%%%%
%%%%%%%%%%%%%%%%%%%%%%%%%%%%%%%%%%%%%%%%%%%%%%%%%%%%%%

\subsection{Universality of Recurrence Time in the Free-Fermionic and Free-Bosonic Regimes}

Recurrence time is proportional to the square of the system size, $L^2$, as shown in eq. (\ref{texperiment}) for all the quantum states  in the free fermionic or the free bosonic regimes of the 1D Bose gas with the delta-function interactions.

In a generic quantum many-body system, the number of eigenstates may be  exponentially large with respect to the particle number.  
We therefore expect that for a generic quantum state 
the recurrence time is proportional to the exponential of an exponential function of the particle number $N$ \cite{Peres}. However, it is not the case in the free fermionic or the free bosonic regime of the 1D Bose gas. 
%\sout{with the delta-function interactions}.

%%%%%%%%%%%%%%%%%%%%%%%%%%%%%%%%%%%%%%%%%%%%%%%%%%%%%%%%%%%
%
% Sec IV
%
\section{Recurrence for Finite Nonzero Values of Interaction Parameter}
%{Recurrence with finite nonzero coupling constants}

\subsection{Superposition of Type-II Excitations}

%%%%%%%%%%%%%%%%%%%%% figure 2 %%%%%%%%%%%%%%%%%%%%%%%%%%%
\begin{figure}[h!]
\includegraphics[width=8cm]{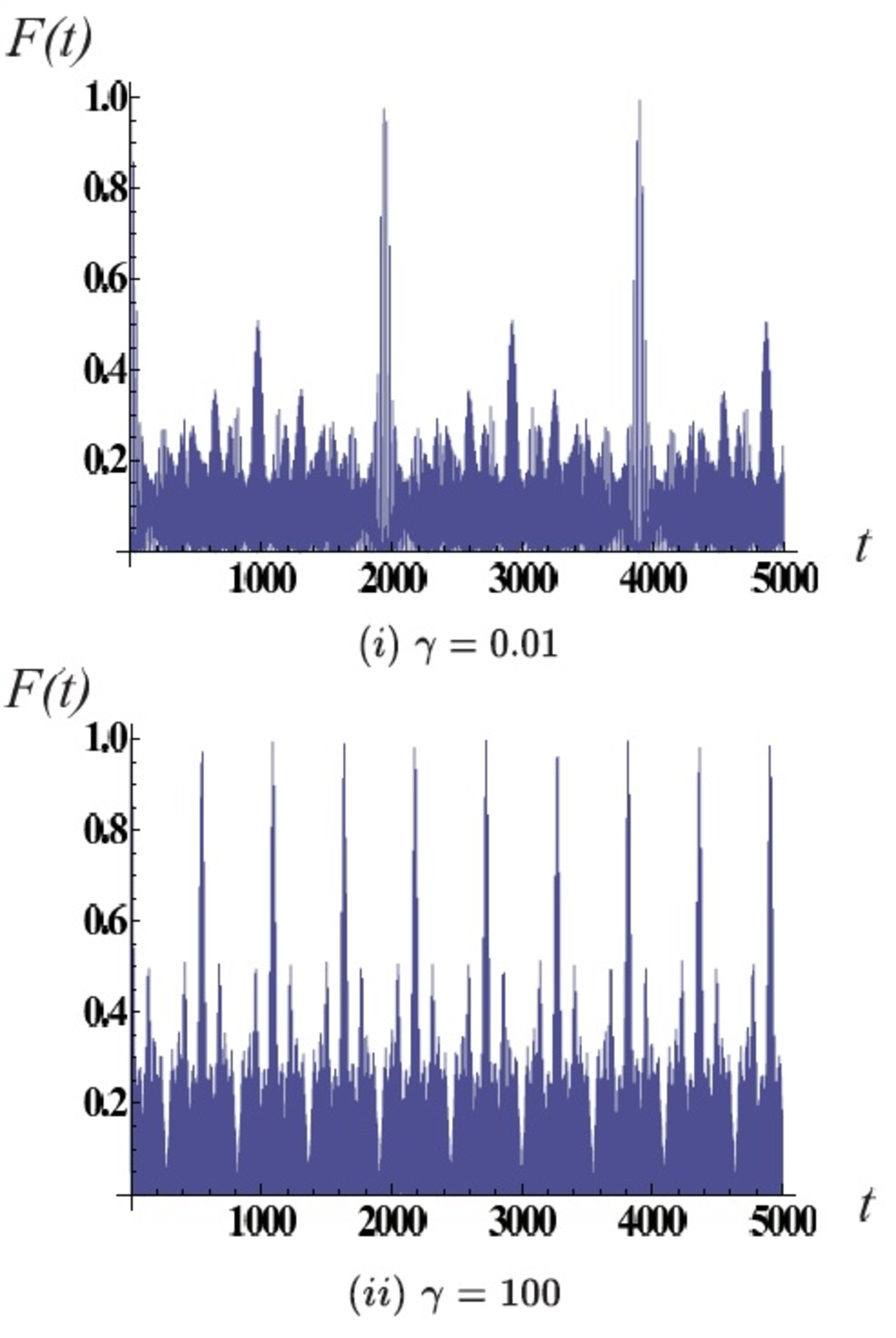}
\caption{
(Color online) Almost complete recurrence: 
Time evolution of the fidelity for the superposition of the type-II excitations 
$F(t) = |\bra X(t) | X(0) \ket |^2 $ 
with $N=12$: 
(i)  $\gamma=0.01$ (upper panel) ; 
(ii)  $\gamma=100$ (lower panel). }
\label{peri3}
\end{figure}
%%%%%%%%%%%%%%%%%%%%%%%%%%%%%%%%%%%%%%%%%%%%%%%%%%%

In the case of finite nonzero values of interaction parameter $\gamma$, 
we calculate the recurrence time for the initial state (\ref{eq:XN}) given by the superposition of one-hole excitations. 
For $\gamma= 0.01$ and 100, the fidelity does not return to 1.0 completely.
However, we sometimes observe that the fidelity becomes close to 1.0 in time evolution.  
Here we recall that we define the recurrence time by the time interval between  the initial time and the time when the fidelity first returns to a value larger than 0.9. 
For example, in the lower panel of Fig. \ref{peri3}, for the first recurrence at $t=538$, 
the value of the fidelity is given by 0.94, where interaction parameter is given by $\gamma=100$ and the number of particles  $N=12$.

\begin{figure}[h]
\includegraphics[width=8cm]{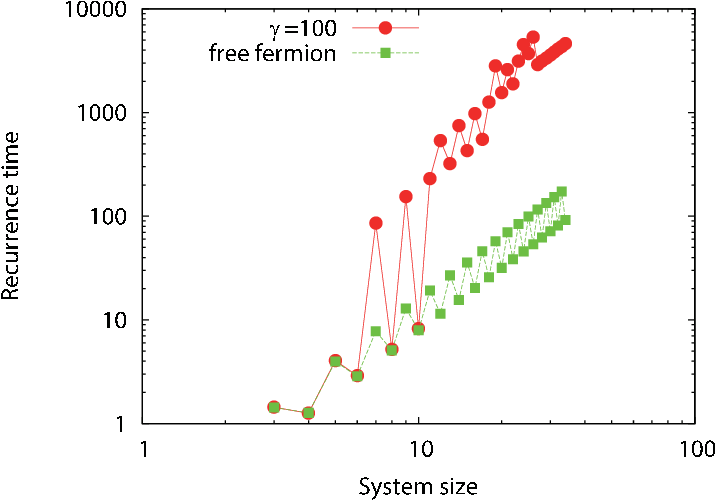}
\vskip 24pt
\caption{(Color online)
Recurrence time $T$ versus system size $N$. 
For the free fermions ($\gamma= \infty$), $T$ is proportional to $N^2$ (green squares). 
For $\gamma=100$,  $T$ is proportional to $N^2$ (red circles), although 
it is 100 times larger than that of the free fermions.  }
\label{c100}
\end{figure}

\begin{figure}[ht]
\includegraphics[width=8cm]{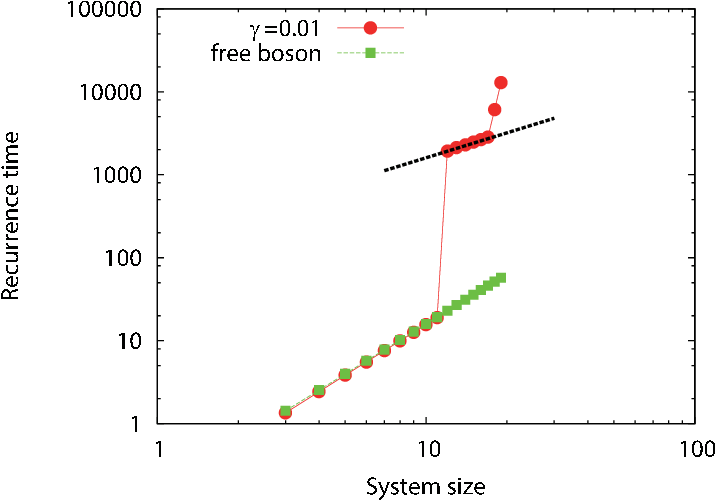}
\vskip 24pt
\caption{ (Color online) 
Recurrence time $T$ versus system size $N$.
For free bosons  ($\gamma=0$) ,  $T$ is proportional to  $N^2$  (green squares).
For $\gamma= 0.01$ $T$ is proportional to $N$  (red filled circles).  
Recurrence time becomes much longer than that of free bosons at $N=12$ as the system size $N$ increases.  
}
\label{c001}
\end{figure}

As interaction parameter  $\gamma$ increases from zero (i.e., in the free-bosonic regime) to a finite non-zero value such as $\gamma=0.01$, which is not extremely small,  
the recurrence time enhances abruptly at some value of $\gamma$. 
We observe that 
the recurrence time is much longer than those of free particles 
as shown in Figs. \ref{c100}  and \ref{c001} for $\gamma= 100$ and $0.01$, respectively.  In Fig. \ref{c100} the recurrence time 
becomes 100 times longer than that of  free-fermions. 
However,  it is still proportional to the square of the system size,  $N^2$ , 
at least up to some value of $N$. 

In Fig. \ref{c001} the recurrence time is almost proportional to the square of the system size,  $N^2$, which is the same as that of the free-bosonic regime if the number of the particle is small such as for $N = 3 \sim 11$, while the recurrence time is proportional to the system size $N$ when the number of the particle is given by $N = 12 \sim 17$. 
For $N \geq 18$, the recurrence time enhances abruptly, and we could not determine it.

Periodic patterns appear in the graph of fidelity $F(t)$ as a function of time. 
In Fig. \ref{peri3} periodic patterns in the time evolution of the fidelity 
are shown in the weak coupling case of  $\gamma=0.01$ (upper panel) 
and in the strong coupling case of  $\gamma=100$ (lower panel) for $N=12$.
The recurrence time for $\gamma=100$ becomes much longer than that in free-fermion, such as 100 times longer. However,  it is still approximately proportional to 
some power of the system size $N$.

%%%%%%%%%%%%%%%%%%%%% figure 3%%%%%%%%%%%%%%%%%%%%%%%%%%%
%%%%%%%%%%%%%%%%%%%%%%%%%%%%%%%%%%%%%%%%%%%%%%%%%%

%%%%%%%%%%%%%%%%%%%%% figure 4 %%%%%%%%%%%%%%%%%%%%%%%%%%%

%%%%%%%%%%%%%%%%%%%%%%%%%%%%%%%%%%%%%%%%%%%%%%%%%%%%%%%%%

%%%%%%%%%%%%%%%%%%%%% figure 5 %%%%%%%%%%%%%%%%%%%%%%%%%%%
\begin{figure}[h]
\includegraphics[width=8cm]{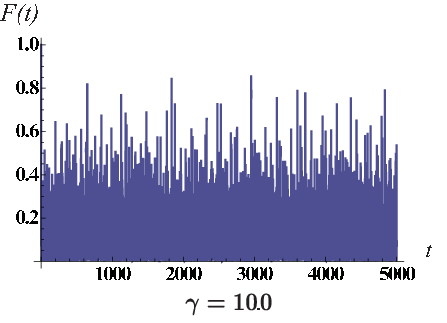}
\caption{ 
(Color online)
Fidelity $F(t) = |\bra X(t) | X(0) \ket |^2 $ shows no periodic patterns 
in time evolution. 
Interaction parameter $\gamma$ is given by $\gamma=10.0$, and 
the particle number is given by $N=10$.}
\label{c10}
\end{figure}
%%%%%%%%%%%%%%%%%%%%%%%%%%%%%%%%%%%%%%%%%%%%%%%%%%
% 

For finite values such as $\gamma= 10$ with $N=10$, the fidelity is always smaller than 0.9 and there is no periodic structure observed in the fidelity as a function of time,  as shown in Fig. \ref{c10}. For these intermediate values of $\gamma$, we do not 
evaluate the recurrence time.

%%%%%%%%%%%%%%%%%%%%%%%%%%%%%%%%%%%%%%%%%%%%%%%%%%%%%%%%%%%
%
%
%
\subsection{Recurrence of the Density Distribution}

We now show the recurrence of the density profile in a quantum state  
and compare it with that of the fidelity.  

Let us introduce the second quantized Hamiltonian of the LL model \cite{Korepin}. 
\begin{align}
{\cal H'}_{\text{LL}} = 
\int_{0}^{L} dx [ \partial_x \hat{ \psi}^{\dagger} \partial_x \hat{ \psi} + 
c \hat{ \psi}^{\dagger} \hat{ \psi}^{\dagger}  \hat{ \psi} \hat{ \psi} ] ,
\end{align}
where $\hat{\psi}(x,t)$ is the canonical Bose field.
We remark that every Bethe eigenvector of the LL model  corresponds to  
an eigenstate of the Hamiltonian ${\cal H'}_{\text{LL}}$ \cite{Korepin}.  
By applying the conjugate field operators to the vacuum state 
with coefficients being given by the wavefunction of the Bethe eigenvector 
we obtain  the corresponding eigenstate of the Hamiltonian ${\cal H'}_{\text{LL}}$. 
We recall that in the LL model every Bethe eigenvector of $N$ particles 
is specified by a corresponding  set of the Bethe quantum numbers $I_j$ for $j=1, 2, \ldots, N$,  and also that $I_j$'s are given by integers if  $N$ is odd and half-odd integers if $N$ is even. Here we remark that it is argued that the Bethe ansatz eigenvectors are complete in the LL model if  the coupling constant $c$ is positive \cite{Dorlas}.   

For a given quantum state $| \Psi \ket$ of $N$ particles,  we assume that it is expressed in terms of the superposition of the Bethe eigenstates $|\{ I_j \}; N \ket$ as 
\begin{align}  
| \Psi \ket = \sum_{ \{I_j \} \in {\cal S} } c_{\{I_j \}} |\{ I_j \}; N \ket \, . 
\end{align} 
Here we denote by ${\cal S}$ the set of sets $\{I _j\}$ of the Bethe quantum numbers $I_j$'s for the Bethe eigenstates.  
We evaluate  the expectation value of the density operator ${\hat \rho}(x, t)={\hat \psi}^{\dagger}(x, t) {\hat \psi}(x, t)$ for the state $|\Psi \ket$ at time $t$ with position $x$ as  
\begin{align}
\bra \Psi | {\hat \rho}(x, t) | & \Psi \ket
=\sum_{ \{I_j \}, \{ I_j^{'} \} \in {\cal S} } c^{*}_{ \{ I_j^{'} \} } c_{ \{ I_j \} } 
 \nn \\ 
&\times e^{i(P-P^{'})x-i(E-E^{'})t} \bra \{ I_j^{'} \}; N | {\hat \rho}(0, 0)  | \{ I_j \}; N \ket,  \label{eq:rho(t)}
\end{align}
where  $P$ and $P'$ ($E$ and $E'$) denote the total momenta (the total energies) of
$|\{I_j \}; N \ket$ and $|\{I^{'}_j \}; N \ket$, respectively.
In eq. (\ref{eq:rho(t)}) we evaluate the form factor 
$\bra\{ I_j^{'} \}; N| {\hat \rho}(0, 0) |\{ I_j \}; N\ket$ 
through the determinant of Slavnov's formula \cite{Slavnov1989}. 
The explicit expression is given in Appendix A.

We denote by $\rho(x, t)$ the expectation value of the density operator at time $t$ and position $x$ given in eq. (\ref{eq:rho(t)})  
\begin{equation} 
\rho(x,t) = \bra \Psi | {\hat \rho}(x, t) |  \Psi \ket .
\end{equation}  
We call the plot of  $\rho(x, t)$  against position $x$  
the density distribution or the density profile at time $t$. 
We thus derive the exact time evolution of the density profile, numerically.  
Once we evaluate the form factors  of the density operator 
at $t=0$ and $x=0$ we obtain the density profile at any later time $t$ only by taking the sum of the exponentials in  Eq. (\ref{eq:rho(t)}).

%%%%%%%%%%%%%%%%%%%%% figure 6 %%%%%%%%%%%%%%%%%%%%%%%%%%%
\begin{figure}[h]
\includegraphics[width=8cm]{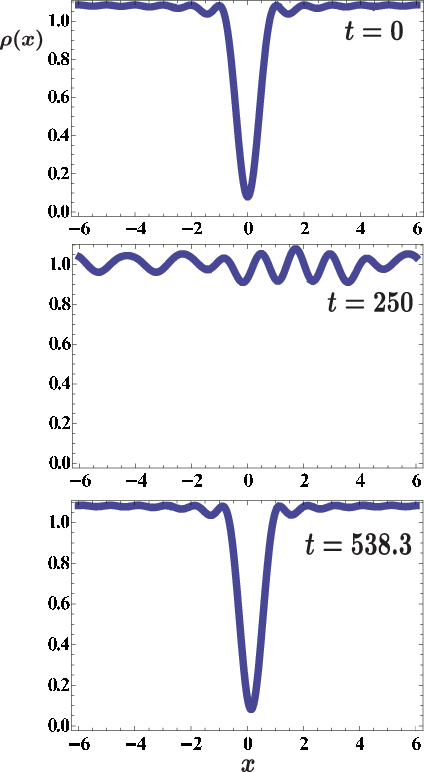}
\caption{ 
(Color online)
Recurrence in density distribution $\rho(x, t)$ for the sum over one-hole excitations 
of $N=12$ and $\gamma=100$. 
%\sout{with $n=N/L=1$.} 
}
\label{density}
\end{figure}
%%%%%%%%%%%%%%%%%%%%%%%%%%%%%%%%%%%%%%%%%%%%%%%%%%

The snapshots  of the density profile $\rho(x, t)$ at three different points of time: $t$=0, 250 and 538.3 are shown  in Fig. \ref{density} for the state given by the sum over one-hole excitations. The initial profile with a localized dip or a density notch once collapses and then returns back to almost the same profile  at the point in time of recurrence, 
i.e. at $t=T$ for recurrence time $T$. 
Here we remark that one can specify the point of time for recurrence quite precisely, since the density profile changes rather quickly within a short period of time around at the recurrence time $t=T$.

\subsection{Recurrence of the Local Density at the Origin}

\begin{figure}[h]
\includegraphics[width=8cm]{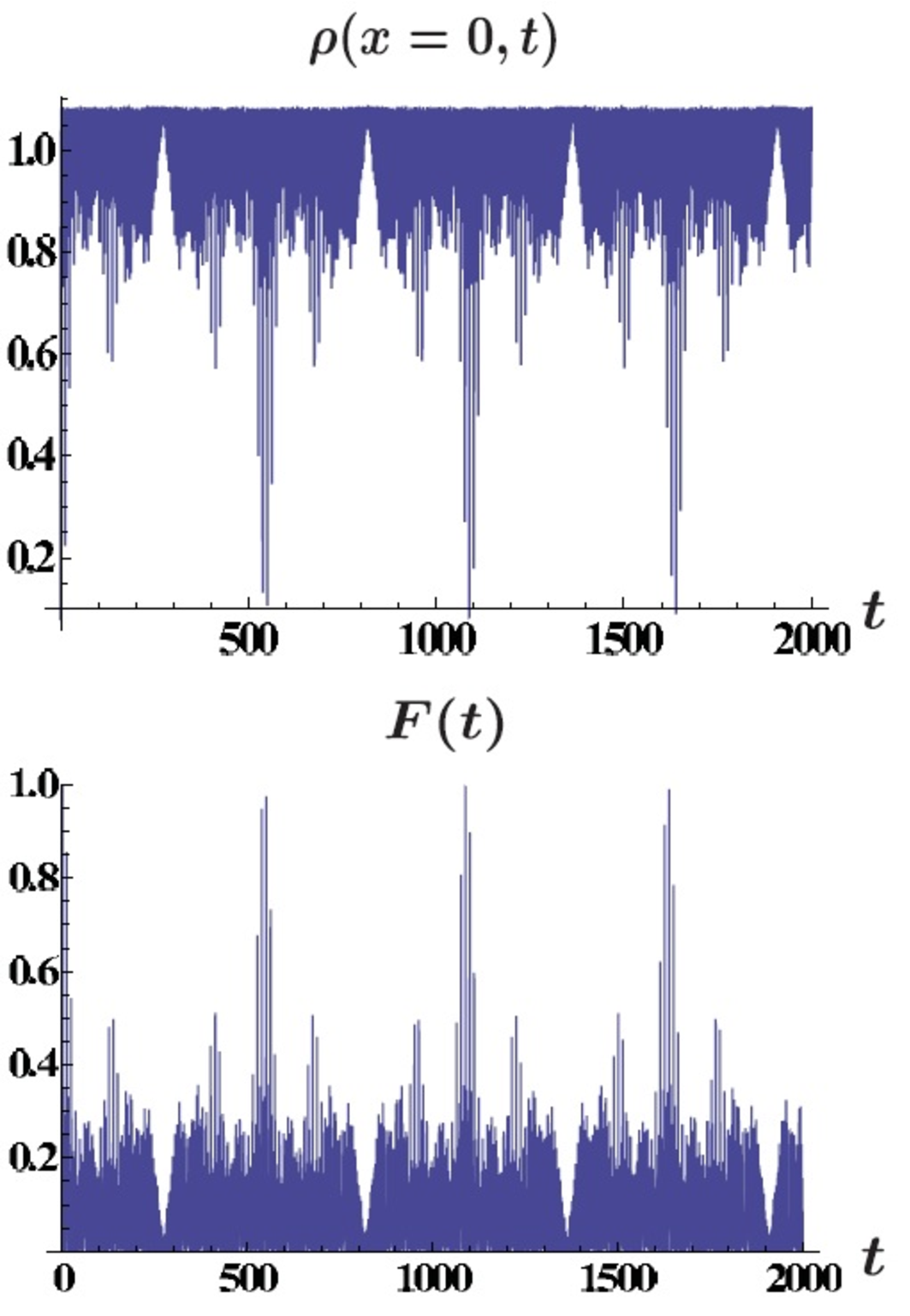}
\caption{ 
(Color online)
Recurrence of the local density at the origin $x=0$, 
$\rho(x=0, t)$, in the upper panel, and that of the fidelity, $F(t)$, 
in the lower panel,  for the quantum state given by the sum over one-hole excitations 
with $N=12$ and $\gamma=100$.
%\sout{with $n=N/L=1$.} 
}
\label{dandf}
\end{figure}

We now compare the time evolution of the local density at $x=0$, i.e. $\rho(x=0, t)$ with that of the fidelity $F(t)$ in the quantum state given by the sum over one-hole excitations for $N=12$.  In Fig. \ref{dandf}  we observe that if the local density at the origin returns to the initial value the fidelity also returns to a value close to 1.0, while 
if otherwise it does not, for the quantum state given by the sum over one-hole excitations with $N=12$.
The observation in Fig. \ref{dandf} is remarkable. It is clear that if the fidelity returns to 1 the density profile also returns to the initial one. However,  the inverse is not always true.

In experiments it is important to know how much information we can obtain about the fidelity if we measure other physical quantities such as the local density at some position. In fact, it seems that it is impossible to measure  directly the fidelity in experiments. 
It follows from the observation of Fig.  \ref{dandf} that in some experiment realizing the 1D Bose gas with the delta-function interactions, if we observe that the local density at a position returns to the initial value,  it practically suggests the recurrence of the whole system at the time when we measure the local density.

%

%%%%%%%%%%%%%%%%%%%%% figure 7 %%%%%%%%%%%%%%%%%%%%%%%%%%%
%%%%%%%%%%%%%%%%%%%%%%%%%%%%%%%%%%%%%%%%%%%%%%%%%%

%%%%%%%%%%%%%%%%%%%%%%%%%%%%%%%%%%%%%%%%%%%%%%%%%%
%
% Sec IV. C
%
\subsection{Recurrence Time for Other Quantum States}

We observe that recurrence occurs for various other types of initial states such as the sum over two-hole excitations, that of one-hole excitations with random weights and that of two-hole excitations with random weights.    
We also observe that periodic patterns appear in the fidelity as a function of time 
for many quantum states such as the sum over two-hole excitations and the sum over one-hole (or two-hole) excitations with random weights. 

%%%%%%%%%%%%%%%%%%%%% figure 8 %%%%%%%%%%%%%%%%%%%%%%
\begin{figure}[h!]
\includegraphics[width=8cm]{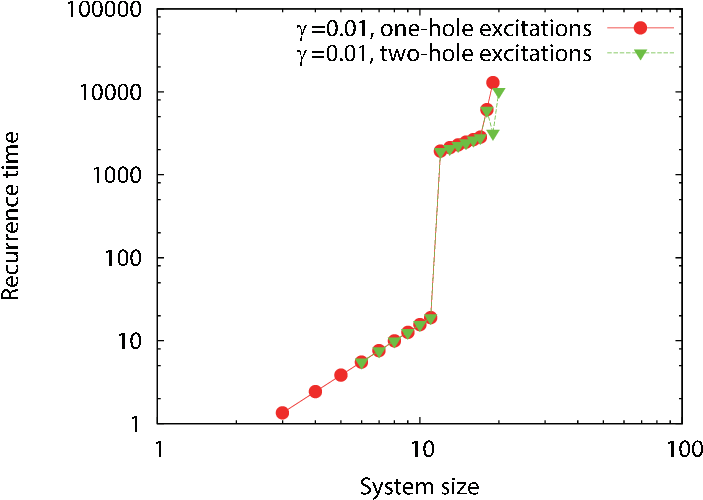}
\vskip 24pt
\caption{ 
(Color online) Recurrence time $T$ versus system size $N$.  
The initial state is constructed by the sum over one-hole excitations (red circles) and 
the sum of two-hole excitations (green lower triangles). 
For $\gamma$ = 0.01, $T$ is approximately
proportional to $N$ 
for $N \ge 12$. 
}
\label{c0012p2h}
\end{figure}
%%%%%%%%%%%%%%%%%%%%%%%%%%%%%%%%%%%%%%%%%%%%%%%%%%
%
%%%%%%%%%%%%%%%%%%%%% figure 9 %%%%%%%%%%%%%%%%%%%%%%
\begin{figure}[h!]
\includegraphics[width=8cm]{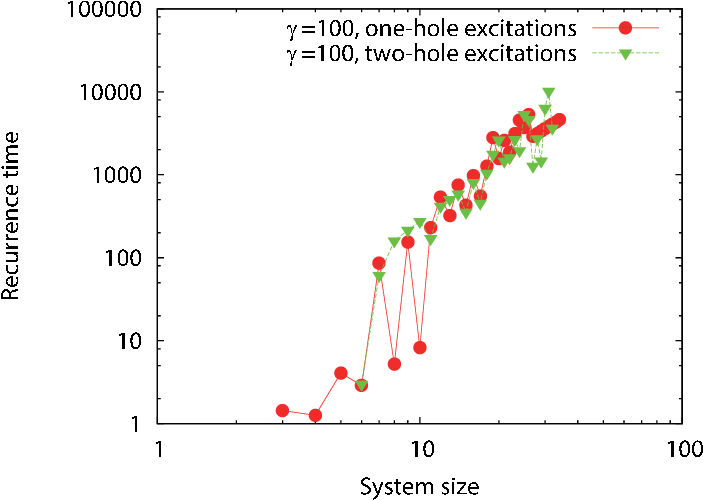} 
\vskip 24pt 
\caption{ (Color online) Recurrence time $T$ versus system size $N$ 
for $\gamma$ = 100. 
Initial state is given by the sum over one-hole excitations (red circles) and the sum of two-hole excitations (green lower triangles). 
For both of them,  $T$ is approximately
proportional to $N^2$.}\label{c1002p2h}\end{figure}
%%%%%%%%%%%%%%%%%%%%%%%%%%%%%%%%%%%%%%%%%%%%%%%%%%

First, we observe that the numerical estimates of recurrence time are 
given by almost  the same values for the two different initial states, 
the sum over one-hole excitations (\ref{eq:XN}) and the sum of two-hole excitations (\ref{x2}), see Fig. \ref{c0012p2h} for $\gamma=0.01$ and Fig. \ref{c1002p2h} for $\gamma=100$, respectively.

In Fig. \ref{c0012p2h} ($\gamma=0.01$) the plots of recurrence time $T$ versus 
system size $N$ overlap each other for the sum of one-hole excitations 
and that of two-hole excitations. 
Thus, as shown in Fig. \ref{c001} the recurrence time becomes 100 times longer than that of  free-fermions at some system size $N$. If $N$ is larger than the number,  
the recurrence time $T$ for the sum of two-hole excitations becomes approximately
proportional to system size $N$ at least up to some large values of $N$.

In Fig. \ref{c1002p2h} ($\gamma=100$) the plots of recurrence time $T$ versus 
system size $N$ overlap each other for the sum of one-hole excitations and that of two-hole excitations. Thus, as shown in Fig. \ref{c100}, the recurrence time $T$ for the sum of two-hole excitations is approximately 
proportional to the square of system size $N$, i.e. $N^2$, at least up to  some value of $N$.  
%

%%%%%%%%%%%%%%%%%%%%%%%%%%%%%%%%%%%%%%%%%%%%%%%%%%%%%%%%%
%
% Sec V
%
\section{Estimation of Recurrence Time in an Experimental System}

Let us estimate the recurrence time in the ultra-cold 1D Bose gas of $^{87}\mbox{Rb}$.
Here we consider the superposition of one-hole excitations 
and make use of   (\ref{tex}) in order to derive the estimate of recurrence time in terms of the dimensionful unit.

We first consider the case of $\gamma=0.01$. 
The recurrence time $\tau_{\bf rec}$ is given by $\tau_{\bf rec} \simeq 4.6 \times 10^8 \times L^2$ [s] up to $N=11$ (see Fig. \ref{c0012p2h}).
For example, $\tau_{\bf rec}  \simeq 460$ [s] for $L=10^{-3}$ [m] and $\tau_{\bf rec} \simeq 46 \times 10^{-3}$ [s] for $L=10^{-5}$  [m].
%, the latter of which is shorter than the experimentally accessible coherence time).
When $N$ exceeds 12, the recurrence time increases very much (see Fig. \ref{c0012p2h}).
For $N=12$, the recurrence time $\tau_{\bf rec}$ is given by $\tau_{\bf rec} \simeq 4 \times 10^{3}$ [s] for $L=10^{-3}$ [m] and  $\tau_{\bf rec} \simeq 0.4$ [s] for $L=10^{-5}$ [m].

Next we consider the case of $\gamma=100$.
Let us first consider odd $N$ cases. When $N=3$ or 5, $\tau_{\bf rec} \simeq 4.6 \times 10^8 \times L^2$ [s] that is, $\tau_{\bf rec} \simeq 500$ [s] for $L=10^{-3}$ [m] and $\tau_{\bf rec} \simeq 50 \times 10^{-3}$ [s] for $L=10^{-5}$  [m].
The recurrence time becomes large for $N \ge 7$ (see Fig. \ref{c1002p2h}):
when $N=7$, for instance, $\tau_{\bf rec} \simeq 5 \times 10^3$ [s] for $L=10^{-3}$ [m] and $\tau_{\bf rec} \simeq 0.5$ [s] for $L=10^{-5}$ [m].
Then, let us consider the case of even $N$. 
From Fig. \ref{c1002p2h}, up to $N=10$, $\tau_{\bf rec} \simeq 2.3 \times 10^{8} \times L^2$ [s]: $\tau_{\bf rec} \simeq 230$ [s] for $L=10^{-3}$ [m] and $\tau_{\bf rec} \simeq 23 \times 10^{-3}$ [s] for $L=10^{-5}$ [m]. 
We recall that when $N \ge 12$, the estimate of recurrence time is approximately  proportional to $N^2$ if it is expressed in terms of the dimensionless unit $t$. 
When $N=12$, $\tau_{\bf rec} \simeq 1 \times 10^4$ [s] for $L=10^{-3}$ [m] and $\tau_{\bf rec} \simeq 1$ [s] for $L=10^{-5}$ [m].

Thus, for $\gamma=0.01$ and 100,  the estimate of recurrence time is 
given by the order of ten milliseconds 
in some  cases of  $L=10^{-5}$[m] and $N$ is up to about 10. 

Finally, we give the estimates of recurrence time 
in the cases of  $\gamma=10^{-3}$ and  $10^{3}$.  
For $L=10^{-3}$ [m] and $N=20$ 
 recurrence time $T$ is given by $T=500$ [s]  at  $\gamma=10^{-3}$
 and by $T=2 \times 10^2$ [s] at $\gamma=10^{3}$.  
 For  $L= 10^{-5}$ [m] and $N=20$, we have $T=50 \times 10^{-3}$ [s] 
 with $\gamma=10^{-3}$, 
 and by $T=20 \times 10^{-3}$ [s] at $\gamma=10^{3}$.

We now suggest that it is an interesting but nontrivial problem how to observe experimentally the relatively short recurrence time in the Bose gas with the delta-function interactions predicted in the paper. It is an isolated quantum many-body system,  and there should be several aspects to be studied. For instance, in order to observe recurrent phenomena experimentally it should be necessary to keep the system being almost completely isolated from the environment. However,  we expect that the  characteristic time for coherence can be taken to be long enough  in  experiments with respect to  the recurrence time.

\section{Conclusion}
We have shown that  the fidelity returns to a value close to 1.0 during time evolution in the quantum many-body system of the 1D Bose gas with the delta-function interactions 
for some initial states such as the sum over one-hole excitations 
and that of two-hole excitations with some finite nonzero values of the interaction parameter $\gamma$ such as $\gamma=0.01$ and 100.    
We have obtained  the estimates of  recurrence time for the initial states.  
We have also shown that the density profile shows recurrence 
for the state given by the sum over one-hole excitations.   
Here we recall that it is very rare to observe a recurrent phenomenon actually in the time evolution of an generic isolated quantum many-body system.

In the free-bosonic and the free-fermionic regimes, we derive the recurrence time rigorously for any given initial state.  It is proportional to $N^2$ in terms of the dimensionless time variable $t$ of eq. (\ref{tex}), while it is proportional to $L^2$ 
in the original unit of time such  as shown in eq. (\ref{texperiment})
with the dimensionful time variable $\tau$.

For the quantum state given by  the sum over one-hole excitations  
in the case of $\gamma$ = 100 the recurrence time is almost proportional to the square of  the system size,  $N^2$,  while in the case of $\gamma$ = 0.01,  the recurrence time is almost proportional to the system size $N$.  
Here we employ the dimensionless time variable $t$.
At some intermediate values of  $\gamma$, such as $\gamma= 10$ for $N=10$, 
there is no periodic structure observed in the fidelity as a function of time  
as shown in Fig. \ref{c10}.  

Finally, the estimate of recurrence time in the 1D Bose gas is given by the order of ten milliseconds in cold atoms confined in one dimension of  ten micrometers in length 
in the original unit of time.

\section*{Acknowledgement}

The authors would like to thank F.~G{\"o}hmann,  R. Kanamoto, A. Kl{\"u}mper,  
T. Monnai and T. Mori for their useful discussions. 
This work was partially supported by the JSPS Institutional Program 
for Young Researcher Overseas Visits, 
and by Grant-in-Aid for Scientific Research No. 24540396.  
E.K. acknowledges support from the JSPS for financial support (Grant No. 2410747) and Institute for Photon Science and Technology.

\setcounter{equation}{0} 
\renewcommand{\theequation}{A.\arabic{equation}}
\setcounter{figure}{0} 
\renewcommand{\thefigure}{A.\arabic{figure}}

\appendix

\section{Determinant Formula of the Form Factors of the Density Operator}

Let us consider the matrix elements of the density operator ${\hat \rho}(0, 0)$  
between the two Bethe eigenstates $| \{ I_j \}; N  \ket$ and $| \{ I^{'}_j; N \ket$. 
We also call it the form factor between the two eigenvectors \cite{Slavnov1989}.  

We evaluate the form factor $\bra \{ I^{'}_j \}; N  | {\hat \rho}(0, 0) | \{ I_j \}; N \ket$ 
in eq. (\ref{eq:rho(t)}) through 
the determinant of Slavnov's formula \cite{Slavnov1989}
\begin{align}
\bra \{ I^{'}_j \}; N  |{\hat \rho}(0, 0)  | \{ I_j \}; N \ket 
=i^N (P-P')&
\(\prod^N_{j,\ell=1}\frac{k_j-k_\ell+ic}{k'_j-k_\ell}\) \nn\\&\times
\det_{N-1}U(k,k'), \label{eq:Slavnov}
\end{align}
where the quasi-momenta $\{k_1,\cdots,k_N\}$ and $\{k'_1,\cdots,k'_N\}$ 
give the eigenstates $| \{ I_j \}; N  \ket$ and $| \{ I^{'}_j; N \ket$, respectively. 
The matrix elements of the $(N-1)$ by $(N-1)$ matrix $U(k,k')$ 
for $j, k= 1, 2, \ldots, N-1$, are given by 
\begin{align}
U(k,k')_{j,\ell}&=2\delta_{j\ell}\text{Im}\[\prod^N_{a=1}
\frac{k'_a-k_j + ic}{k_a-k_j + ic}\]
+\frac{\prod^N_{a=1}(k'_a-k_j)}{\prod^N_{a\neq j}(k_a-k_j)} \nn\\&\times
\(K(k_j-k_\ell)-K(k_N-k_\ell)\), \label{eq:matrixU}
\end{align}
where $K(k)=2c/(k^2+c^2)$.

Thus,  the evaluation of the form factors of the Bethe eigenstates with $N$ particles are reduced to that of the determinants (\ref{eq:matrixU})  by making use of eq. (\eqref{eq:Slavnov}).

\end{document}